\documentclass[prl,aps,reprint,showpacs,amssymb,superscriptaddress]{revtex4-1}
\usepackage{graphicx}
\usepackage{dcolumn}
\usepackage{bm}

\newcommand{\hcd}{\hat{c}^{\dagger}}

\newcommand{\hc}{\hat{c}^{\phantom{\dagger}}}
\begin{document}
\title{Antiferromagnetic order in multi-band Hubbard models for
iron-pnictides}
\author{T.~Schickling}
\affiliation{Fachbereich Physik, Philipps Universit\"at, Renthof 6,
35032 Marburg, Germany}
\author{F.~Gebhard}
\affiliation{Fachbereich Physik, Philipps Universit\"at, Renthof 6,
35032 Marburg, Germany}
\author{J.~B\"unemann}
\affiliation{Fachbereich Physik, Philipps Universit\"at, Renthof 6,
35032 Marburg, Germany}
\affiliation{Institut f\"ur Physik, BTU Cottbus, P.O. Box 101344, 
03013 Cottbus, Germany}

\date{\today}
\begin{abstract}
We investigate multi-band Hubbard models for
the three iron 3$d$-$t_{2g}$ bands 
and the two iron 3$d$-$e_g$ bands in ${\rm La O Fe As}$ 
by means of the Gutzwiller variational theory. 
Our analysis of the paramagnetic ground 
state shows that neither Hartree--Fock mean-field theories 
nor effective spin models describe these systems adequately.
In contrast to Hartree--Fock-type approaches, 
the Gutzwiller theory predicts that antiferromagnetic order 
requires substantial values of the local Hund's-rule
exchange interaction. For the three-band model,
the antiferromagnetic moment fits experimental data 
for a broad range of interaction parameters. However, for the
more appropriate five-band model, the iron $e_g$ electrons
polarize the $t_{2g}$ electrons and they
substantially contribute to the ordered moment.
\end{abstract}

\maketitle

Since their recent discovery, the iron-based high-$T_{\rm c}$ 
superconductors have attracted tremendous attention both in
theory and experiment. From a theoretical point of view,
these systems are of particular interest because
their conduction electrons are less correlated 
than those of other high-$T_{\rm c}$ superconductors.
In contrast to the cuprates, the pnictides' undoped parent compounds
are antiferromagnetic metals at low temperatures,
not insulators. However, the electronic mass is enhanced by
a factor of two which indicates that electronic correlations
are quite substantial in the pnictides, too.

The theoretical description of the pnictides' normal phase turned out to
be a difficult problem. Standard density-functional theory (DFT)
grossly overestimates the size of their magnetic moment
in the antiferromagnetic ground state.
For example, in ${\rm La O Fe As}$ experiment finds 
a staggered moment of 
$m= (0.4\ldots 0.8)\mu_{\rm B}$~\cite{magmom1,magmom2,magmom3} 
whereas DFT calculations predict moments of 
$m\approx 1.8\mu_{\rm B}$ or larger~\cite{magmomdft1,magmomdft2}. 
For other pnictide compounds, the comparison is equally unfavorable.

Perturbative many-electron theories for the pnictides face the problem 
that all five iron 3$d$-bands contribute significantly to the band
structure near the Fermi energy. Even when the effective 
Hubbard interaction~$U$ between pairs of electrons on the same iron atom
is smaller than the effective 3$d$~bandwidth~$W$, $U\lesssim W/2$,
atomic charge fluctuations with ionization degree $\Delta N>U/W$
are improbable in the ground state. The description
of itinerant antiferromagnetism in such correlated multi-band systems
is beyond perturbative mean-field approximations such as
Hartree--Fock (HF) theory. 
In studies based on dynamical mean-field theory (DMFT),
only the paramagnetic phases of these models could be 
investigated so far, see e.g., Ref.~\cite{ishida2010}.  

In this work, we employ the Gutzwiller variational theory (GT) which describes
the ground state and quasi-particle excitations of multi-band Fermi liquids;
for an application to nickel, see Refs.~\cite{buenemann2003,hofmann2009}.
As compared to DMFT, our method is numerically much less costly.
This enables us to resolve the small energy differences 
between the paramagnetic and antiferromagnetic phases 
in the pnictides (N\'eel temperature $T_{\rm N}\approx 140\, {\rm K}$). From 
our study of a five-band model
for the iron 3$d$-$t_{2g}$ and 3$d$-$e_g$ bands
we find that the paramagnetic phase is much more stable than predicted by DFT 
or HF calculations. Only for large values of the 
local exchange interaction, an antiferromagnetic 
ground state is found; in a related study~\cite{chinesen},  
only paramagnetic states were examined.

We investigate the two-dimensional five-band Hubbard model
\begin{equation}
\hat{H}=\sum_{i,j;b,b';\sigma}t_{i,j}^{b,b'}                 
\hat{c}_{i,b,\sigma}^{\dagger}\hat{c}_{j,b',\sigma}^{\phantom{\dagger}}                 
+\sum_i \hat{H}_{{\rm C},i}\equiv\hat{H}_0+\hat{H}_{\rm loc} \;,
\label{1.1} 
\end{equation}       
where $\hat{H}_0$ describes the transfer of electrons with 
spin $\sigma=\uparrow,\downarrow$ between iron atoms on lattice sites $i,j$ 
in the 3$d$-($e_g,t_{2g}$) orbitals $b,b'$. The transfer
parameters $t_{i,j}^{b,b'}$ are taken from Ref.~\cite{graser2009}. 
The bare bandwidth of the 3$d$ electrons is $W=4.8\, {\rm eV}$,
and there are, on average, six electrons on every iron atom.
The local Hamiltonian $\hat{H}_{{\rm C},i}$ describes the Coulomb 
interaction of the iron electrons.
Frequently the following form for the Hubbard interaction is employed,
{\arraycolsep=1pt\begin{eqnarray}
\hat{H}^{(1)}_{\rm C}&=&\hat{H}^{\rm dens}_{\rm C}+\hat{H}^{\rm sf}_{\rm C} \; ,
\nonumber\\[3pt]
\hat{H}^{\rm dens}_{\rm C}&=&
\sum_{b,\sigma}U(b,b)
\hat{n}_{b,\sigma}\hat{n}_{b,\bar{\sigma}}
+\!\!\!\sum_{{\scriptstyle b(\neq)b'}\atop 
{\scriptstyle \sigma,\sigma'}}\widetilde{U}_{\sigma,\sigma'}(b,b')
\hat{n}_{b,\sigma}\hat{n}_{b',\sigma'} \, ,\nonumber\\
\hat{H}^{\rm sf}_{\rm C} &=&
\sum_{b(\neq)b'}J(b,b')\left(\hcd_{b,\uparrow}\hcd_{b,\downarrow}
\hc_{b',\downarrow}\hc_{b',\uparrow}+ {\rm h.c.}\right) \label{app3.5}\\
&&+\sum_{b(\neq)b';\sigma}J(b,b')\hcd_{b,\sigma}\hcd_{b',\bar{\sigma}}
\hc_{b,\bar{\sigma}}\hc_{b',\sigma}\;.
\nonumber
\end{eqnarray}}Here, we dropped the lattice-site indices and 
introduced the abbreviations $\bar{\uparrow}=\downarrow$, 
$\bar{\downarrow}=\uparrow$, and
$\widetilde{U}_{\sigma,\sigma'}(b,b')= U(b,b')-\delta_{\sigma,\sigma'}J(b,b')$,
where $U(b,b')$ and  $J(b,b')$ are the local Coulomb and exchange 
interactions. 

Even in cubic symmetry, however, the Hamiltonian~(\ref{app3.5})
is incomplete. The full Hamiltonian reads 
$\hat{H}_{\rm C}=\hat{H}^{(1)}_{\rm C}+\hat{H}^{(2)}_{\rm C}$ with
{\arraycolsep=1pt\begin{eqnarray}
\hat{H}^{(2)}_{\rm C}&=&\bigg[\sum_{t; \sigma,\sigma'}(
T(t)-\delta_{\sigma,\sigma'}A(t))
\hat{n}_{t,\sigma}\hcd_{u,\sigma'}\hc_{v,\sigma'}
\nonumber\\
&&+\sum_{t,\sigma}A(t)
\left(
\hcd_{t,\sigma}\hcd_{t,\bar{\sigma}}
\hc_{u,\bar{\sigma}}\hc_{v,\sigma}+
\hcd_{t,\sigma}\hcd_{u,\bar{\sigma}}
\hc_{t,\bar{\sigma}}\hc_{v,\sigma}
\right)
\label{h2}\\
&&+\!\!\sum_{{\scriptstyle t(\neq)t'(\neq)t^{\prime \prime}}\atop 
{\scriptstyle e, \sigma,\sigma'}}
\!S(t,t';t^{\prime \prime},e)
\hcd_{t,\sigma}\hcd_{t',\sigma'}
\hc_{t^{\prime \prime},\sigma'}\hc_{e,\sigma}\bigg ]+{\rm h.c.}\,.
\nonumber
\end{eqnarray}}Here, $t$ and $e$ are indices for the  
three $t_{2g}$ orbitals with symmetries $xy$, $xz$, and $yz$,
and the two $e_g$ orbitals with symmetries
$u=3z^2-r^2$ and $v=x^2-y^2$. 
The parameters in~(\ref{h2}) are of the same order of magnitude 
as the exchange interactions $J(b,b')$ and, hence, 
there is no a-priori reason to neglect them. 
Of all the parameters $U(b,b')$, $J(b,b')$,
$A(t)$, $T(t)$, $S(t,t';t^{\prime \prime},e)$ 
only ten are independent in cubic symmetry. 
In a spherical approximation, they are all determined, e.g., by the 
three Racah parameters $A,B,C$. For details on the multiplet structure 
of  $d$-shells we refer to Sugano's textbook~\cite{sugano1970}. 
 
In this work, we work with the orbital averages 
$J\propto\sum_{b\neq b'}J(b,b')$, 
and $U'\propto\sum_{b\neq b'}U(b,b')$ 
of the exchange and the inter-orbital Coulomb interaction. 
They are related to the intra-orbital interaction 
$U=U(b,b)$ via $U'=U-2J$ . 
Due to this symmetry relation, the three values of $U,U',$ and $J$ do not 
determine the Racah parameters $A,B,C$ uniquely. Therefore, we further 
assume that the approximate atomic relation $C/B=4$ is satisfied in solids, too.
In this way, the three Racah parameters and, consequently, all parameters 
in $\hat{H}_{\rm C}$ are functions of $U$ and~$J$.
This permits a meaningful comparison of our results with
those of previous work where, as an additional approximation, 
the parameters $U(b,b')$,  $J(b,b')$ were assumed 
to be orbital independent, see, e.g., Ref.~\cite{ishida2010}. 
We note that the Gutzwiller method is applicable to 
the general form~(\ref{1.1}) of the atomic Hamiltonian.

We approximate the true ground state of $\hat{H}$ in~(\ref{1.1})
by the Gutzwiller variational wave function 
\begin{equation}
|\Psi_{\rm G}\rangle=\hat{P}_{\rm G}|\Psi_0\rangle
=\prod_{i}\hat{P}_{i}|\Psi_0\rangle\;,
\label{1.3} 
\end{equation}
where $|\Psi_0\rangle$ is a single-particle product state, 
and the local `Gutzwiller correlator' is defined as
\begin{equation}
\hat{P}_{i}=\sum_{\Gamma}\lambda_{\Gamma}
|\Gamma \rangle_{i} {}_{i}\langle \Gamma |\;.
\label{1.4}
 \end{equation}
 Here, we introduced variational parameters $\lambda_{\Gamma}$ 
for each of the atomic multiplet states $|\Gamma \rangle_{i}$, 
i.e., the eigenstates of $\hat{H}_{{\rm C},i}$. 
Note that the single-particle state $|\Psi_0\rangle$
is also a variational object which we determine from the minimization 
of the variational energy functional that results from 
the wave functions~(\ref{1.3}). 
The Hartree--Fock theory is a special case of the Gutzwiller theory
which we obtain by setting $\lambda_{\Gamma}=1$ 
for all $|\Gamma \rangle$. 

The evaluation of expectation values with respect to~(\ref{1.3}) 
poses a difficult many-particle problem. 
As shown in Refs.~\cite{buenemann1998,buenemann2005}, 
it can be solved exactly in the limit of infinite spatial dimensions. 
The analytic energy functional derived in this limit can be used 
as an approximation for finite dimensional systems. 
On top of the approximate ground-state description 
which is provided by the Gutzwiller wave function one can also calculate 
quasi-particle band structures  for a comparison with data from
angle-resolved photoemission spectroscopy (ARPES)~\cite{buenemann2003b}.  

The Gutzwiller energy functional contains two different sets of variational 
parameters. The local multiplet occupations are governed by the 
parameters $\lambda_{\Gamma}$ in the correlation operator~(\ref{1.4}). 
As shown in Refs.~\cite{buenemann2005,buenemann2003b}, the single-particle wave
function  $|\Psi_0\rangle$ is the ground state of an effective 
single-particle Hamiltonian 
\begin{equation}
\hat{H}_0^{\rm eff}=\sum_{i,j;b,b';\sigma}\widetilde{t}_{i,j}^{b,b'}             
\hat{c}_{i,b,\sigma}^{\dagger}\hat{c}_{j,b',\sigma}^{\phantom{\dagger}} + 
  \sum_{i;b,b';\sigma} \eta^{b,b'}_{i,\sigma} 
 \hat{c}_{i,b,\sigma}^{\dagger}\hat{c}_{i,b',\sigma}^{\phantom{\dagger}} 
\end{equation}
with renormalized electron transfer parameters $\widetilde{t}_{i,j}^{b,b'}$ 
and variational parameters  $\eta^{b,b'}_{i,\sigma}$ which govern the  
orbital and spin dependent local densities. 
There are ${\cal N}_{\lambda}=2^{10}=1024$ parameters $\lambda_{\Gamma}$ 
while the number of parameters $\eta^{b,b'}_{i,\sigma}$ is much smaller,
${\cal N}_{\eta,{\rm p}}=5$ for the paramagnetic case
and ${\cal N}_{\eta,{\rm af}}=10$ for the antiferromagnetic case.
Nevertheless, the minimization of the energy functional 
with respect to the parameters $ \eta^{b,b'}_{i,\sigma}$ is numerically
expensive because any change of these parameters results in 
a minimization cycle with a full momentum-space integration.
In order to obtain 
the required energy resolution, these integrals have been calculated 
on a momentum-space grid with up to $3\cdot 10^5$  
triangles in the Brillouin zone.

\begin{figure}[tbp]
\begin{minipage}{8.5cm}
\includegraphics[height=5cm]{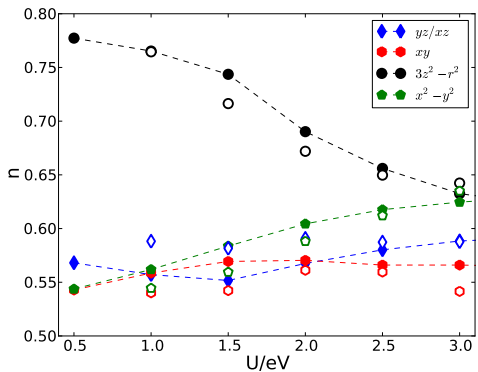}
\end{minipage}
\caption{(Color online) Orbital densities in GT (full symbols)
and in DMFT (open symbols)
as a function of $U$ (with $U/J=4$) for the simplified local Hamiltonian
$\hat{H}_{\rm C}=\hat{H}_{\rm C}^{(1)}$, see~(\protect\ref{app3.5}).\label{fig1}}
\end{figure}

In order to test the reliability of our approach for the five-band model, 
we first compare our results for the partial densities
with those from paramagnetic DMFT calculations. In Fig.~\ref{fig1} 
we show the density of electrons in each orbital 
as a function of~$U$ for fixed ratio  $U/J=4$.
The full symbols give the GT result 
for the simplified local Hamiltonian~(\ref{app3.5}),
$\hat{H}=\hat{H}_0+\hat{H}_{\rm C}^{(1)}$;
open symbols give the DMFT results.
Obviously, the agreement between the GT and DMFT is very good
despite the fact that the particle number is not perfectly conserved
in the DMFT calculations~\cite{note2}.

Fig.~\ref{fig1} shows a common feature of multi-band model systems.
The local Coulomb interaction induces a substantial charge flow between
the bands because, for the local Coulomb interaction, it is energetically
more favorable to distribute electrons equally among the bands. However,
the bands described by $\hat{H}_0$ are usually extracted from a DFT calculation
whose predictions for the Fermi surface reproduce experimental data
reasonably well. Therefore, we argue that the artificial charge flow 
as seen in Fig.~\ref{fig1} is a consequence
of the double counting of Coulomb interactions.
Since the (paramagnetic) Fermi surface found in DFT
reproduces its experimentally determined shape, 
we assume that the same holds for the paramagnetic 
orbital densities. For each value of the interaction parameters we therefore 
choose orbital on-site energies $\epsilon_b=t_{i,i}^{b,b}$ 
which lead to a paramagnetic ground state with the same orbital densities 
as in DFT.  
      
\begin{figure}[htbp]
\begin{minipage}{8.5cm}
\begin{tabular}{@{}ll@{}}
\includegraphics[width=4.1cm]{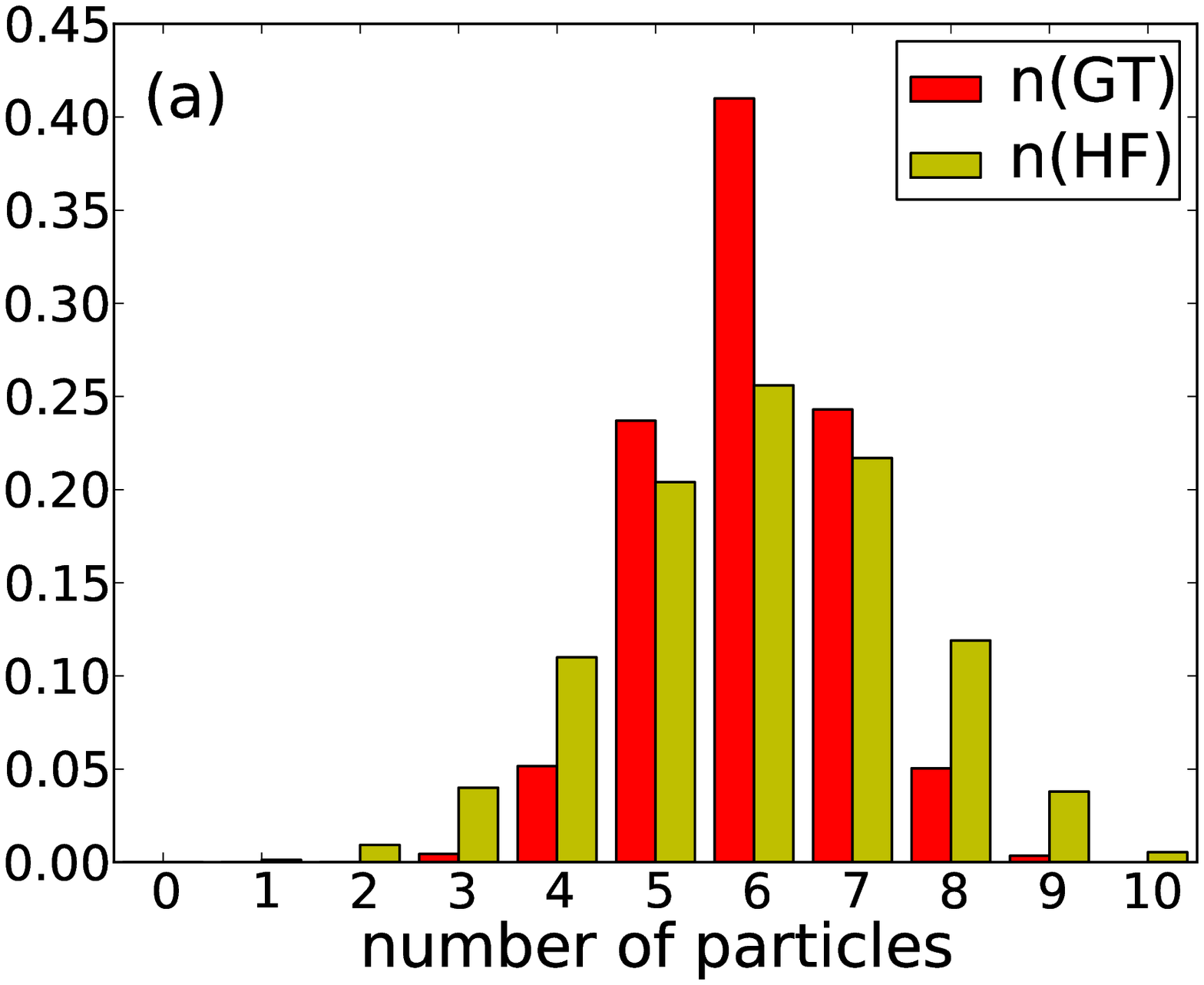} &
\includegraphics[width=4.1cm]{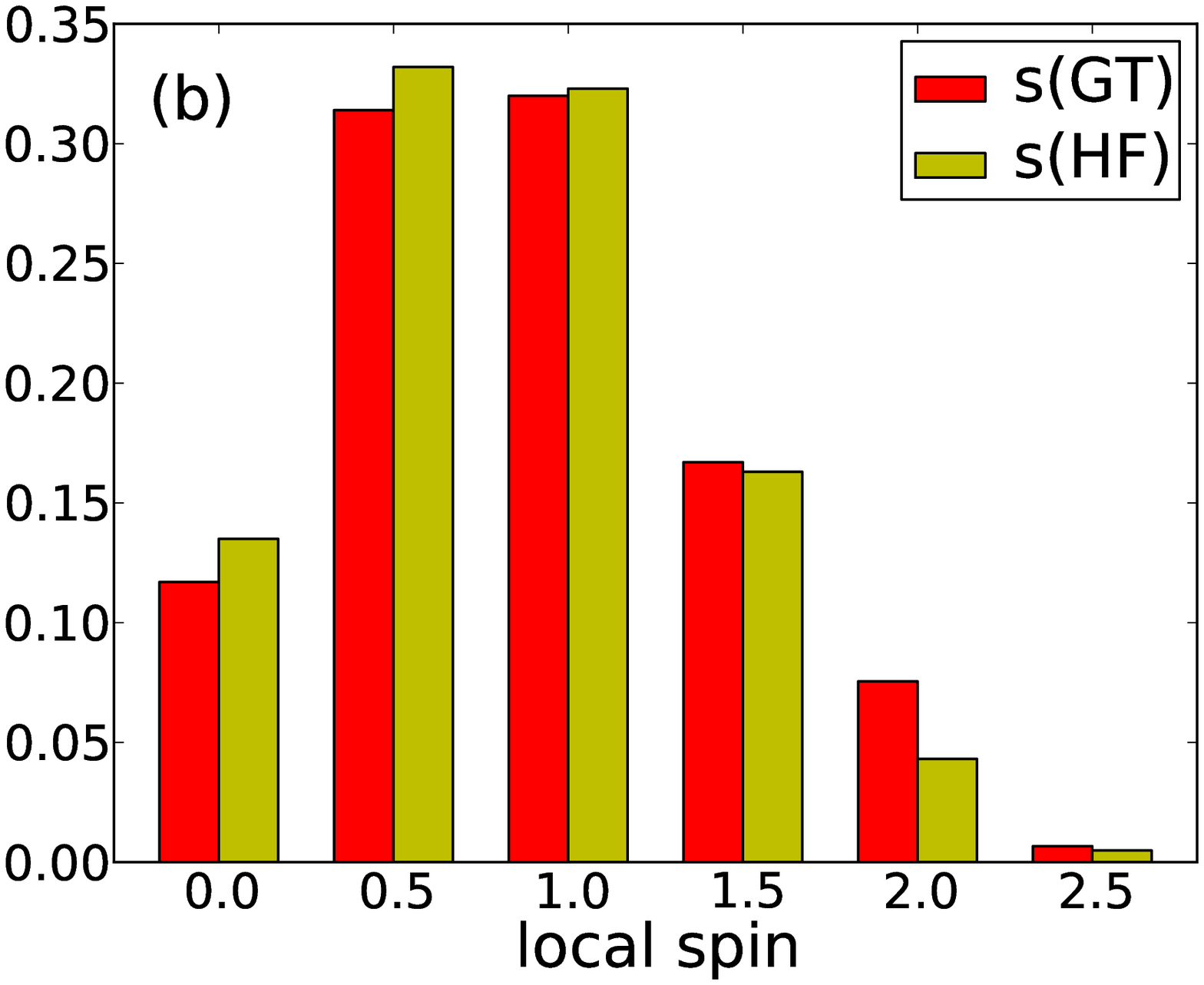}
\end{tabular}
\end{minipage}
\caption{(Color online) Local charge distribution (a) and
spin distribution (b) for the paramagnetic
optimized Gutzwiller wave function for $U=2.6\,$eV, $J=0.4\,$eV 
(left columns) and for the Hartree--Fock wave function 
(right columns).\label{fig2}}
\end{figure}

Next, we investigate the local charge distribution and 
spin distribution in the paramagnet.
In Fig.~\ref{fig2}a we show the probabilities 
to find an atom with $0\leq n\leq 10$ electrons
in $|\Psi_{\rm G}^{\rm opt}\rangle$ for $U=2.6\, {\rm eV}$ 
and $J=0.4\, {\rm eV}$.
As seen from Fig.~\ref{fig2}a, local charge 
fluctuations with $\Delta N = |N-6|>2$ are essentially forbidden. 
Hartree--Fock-type mean-field approximations 
cannot describe this effect properly.
The probabilities to find the local spins $0\leq s\leq 5/2$ 
in $|\Psi_{\rm G}^{\rm opt}\rangle$ are shown in in Fig.~\ref{fig2}b.
The broad spin distribution in Gutzwiller theory is very similar 
in Hartree--Fock theory and shows that the system is far 
from the local-moment situation 
where we would solely find atoms with Hund's-rule spin $s=2$. Consequently,
the elementary excitations of the five-band Hamiltonian~(\ref{1.1})
with parameters from Ref.~\cite{graser2009} 
and ($U=2.6\, {\rm eV}$, $J=0.4\, {\rm eV}$)
are not well approximated by an effective spin model.

\begin{figure}[tbp]
\includegraphics[height=5cm]{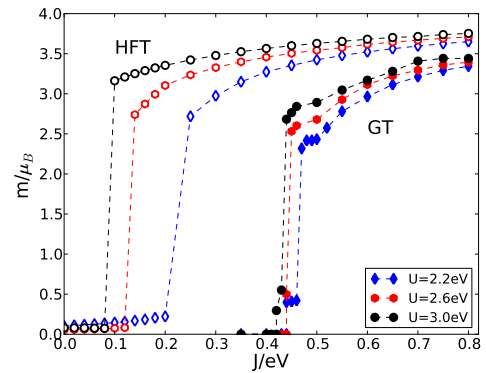}
\caption{(Color online) Magnetic moments in the five-band model
in Hartree--Fock (HFT) and Gutzwiller theory (GT) as a function 
of $J$ for $U=2.2\, {\rm eV}, 2.6\, {\rm eV}, 3.0\, {\rm eV}$.\label{fig4}}
\end{figure}

We now turn to the magnetic properties of our five-band model~(\ref{1.1}).
As seen in Fig.~\ref{fig4}, Hartree--Fock theory predicts
an antiferromagnetic phase for moderate 
$2.2\, {\rm eV}\leq U\leq 3.0\, {\rm eV}$
for all $J>0$.
The HF moments are strongly orbital dependent.
For small values of~$J$, these moment are anti-aligned
which leads to a phase with a small total moment. At critical values
$J=J_c(U)$, the HF moments align which results in a large-moment phase.
Similar phases have been reported in LDA+U calculations~\cite{lda+u}. 

Our Gutzwiller theory shows that 
the correlated paramagnetic state is stable over a wide range of 
Coulomb and exchange-interaction parameters, e.g.,
$J\lesssim 0.4\, {\rm eV}$ for 
$2.2\, {\rm eV}\leq U\leq 3.0\, {\rm eV}$ in Fig.~\ref{fig4}.
The small-$J$ phase with
orbital-dependent moments as seen in Hartree--Fock theory is absent.
The robustness of the paramagnetic phase against 
symmetry breaking is in agreement with earlier 
findings~\cite{buenemann2007b} that Hartree--Fock theory 
generally overrates the importance of
orbitally ordered phases in the ground state of multi-band systems. 

In Gutzwiller theory, the transition to an antiferromagnet with a large moment,
$m\gtrsim 2\mu_{\rm B}$, is as abrupt as in Hartree--Fock theory.
As seen in Fig.~\ref{fig4},
we find an antiferromagnetic state with a small moment $m=0.4\mu_{\rm B}$ 
only in a small region of parameter space, e.g., $U=2.2\, {\rm eV}$ and 
$0.42\, {\rm eV}<J<0.45\, {\rm eV}$. 
Since this region in the $U$-$J$ parameter space 
is fairly small, we do not consider it very likely that the iron
pnictides fall into this parameter region. The explanation for their
small antiferromagnetic moment should have a more natural explanation.

We have extended our analysis to a broader class of variational wave functions
where we included the mixing of atomic configurations in the Gutzwiller
correlator~(\ref{1.4}) of the form $\lambda_{\Gamma,\Gamma'}
|\Gamma \rangle_{i} {}_{i}\langle \Gamma' |$ for $\Gamma\neq \Gamma'$.
In order to keep the problem numerically tractable,
we considered the dominant 500 couplings for which
$|\langle \Psi_{\rm G}^{\rm opt}|\Gamma\rangle_{i}
{}_{i}\langle \Gamma'|\Psi_{\rm G}^{\rm opt}\rangle|^2$ 
is largest in the paramagnetic Gutzwiller state $|\Psi_{\rm G}^{\rm opt}\rangle$.
The inclusion of these additional variational parameters lowered the
energies of the paramagnetic and the antiferromagnetic optimal states
by almost equal amounts so that the phase diagram does not change
noticeably. In particular, the region in phase space with a low magnetic moment
increases only marginally.

\begin{figure}[htbp]
\includegraphics[height=5cm]{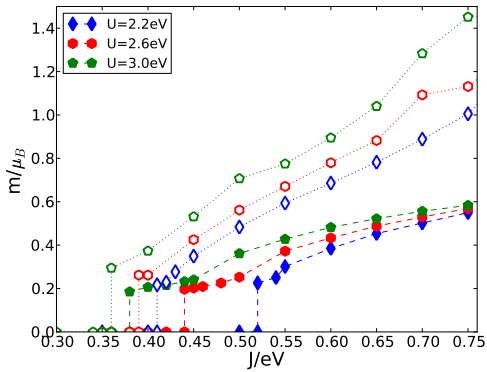}
\caption{(Color online) Magnetic moments for the three-band model
as a function of~$J$ for 
$U=2.2\, {\rm eV}, 2.6\, {\rm eV}, 3.0\, {\rm eV}$.
Full symbols: GT with the full local interaction 
$\hat{H}_{\rm C}=\hat{H}_{\rm C}^{(1)}$~(\protect\ref{app3.5});
open symbols: GT with density-interactions only,
$\hat{H}_{\rm C}=\hat{H}_{\rm C}^{\rm dens}$, cf.~\protect\cite{zhou2010}.
\label{fig3}}
\end{figure}

Our findings for the five-band model contrast those of a recent study of a
three-band model for the pnictides~\cite{zhou2010} where
the authors report a small magnetic moment 
over a large region of Coulomb and exchange interactions. 
In that study, the two 3$d$-$e_g$-bands 
and their interactions $H_{\rm C}^{(2)}$ with
the 3$d$-$t_{2g}$-bands, eq.~(\ref{h2}), were not considered. Moreover,
the spin-flip and pair-exchange terms in the 
local Hamiltonian, the terms $H_{\rm C}^{\rm sf}$ in eq.~(\ref{app3.5}),
have been neglected. 

In Fig.~\ref{fig3} we show the Gutzwiller
prediction for the magnetic moments of the three-band model 
for $2.2\, {\rm eV}\leq U\leq 3.0\, {\rm eV}$
with the full local interaction, $\hat{H}_{\rm C}=\hat{H}_{\rm C}^{\rm dens}+
\hat{H}_{\rm C}^{\rm sf}$, and, for comparison, the result
with $\hat{H}_{\rm C}=\hat{H}_{\rm C}^{\rm dens}$.
Note that we used the electron transfer amplitudes of
Ref.~\cite{zhou2010} for the parameterization of $\hat{H}_0$ in~(\ref{1.1}),
and that we use a slightly different minimization algorithm.
In contrast to the full five-band model,
a large region with a small magnetic moment is found for the three-band model.
When the spin-flip and pair-transfer terms $\hat{H}_{\rm C}^{\rm sf}$
are included in the local interaction, 
the transition in the three-band model
occurs at larger values of~$J$ and, for given $J>J_c$,
the magnetic moment is generally smaller for the full local
interaction than for density interactions only. 
We obtain qualitatively similar results 
when we switch off the coupling between the $t_{2g}$~bands 
and the $e_g$~bands in $\hat{H}_{\rm C}$ of our five-band model
with parameters for $\hat{H}_0$ taken from Ref.~\cite{graser2009}
(`3+2'-band model'). 

The comparison of the results for the three-band model and the five-band model
shows that the Coulomb coupling between the 3$d$-$t_{2g}$-bands
and the 3$d$-$e_g$-bands is very important.
In the five-band model the 3$d$-$e_g$-bands 
contribute substantially to the magnetic
moment and they strongly spin polarize the 3$d$-$t_{2g}$ bands.
In the three-band model (or the `3+2-band model'),
the 3$d$-$e_g$ electrons are essentially paramagnetic. Since the predictions
of the three-band model fit experiment on the magnetic moment of the pnictides
whereas the five-band model fails, it remains to justify the assumption
that the 3$d$-$e_g$ electrons are irrelevant for the magnetism
of the pnictides. In principle, the effective Coulomb parameters
which couple the $t_{2g}$-electrons and the $e_g$-electrons
could be much smaller than those used in our study. In such a
`3+2-band model', the 3$d$-$e_g$ electrons 
would not carry a magnetic moment or polarize the 3$d$-$t_{2g}$-electrons.
More important for the pnictides is the hybridization 
of the irons' 3$d$-electrons with the arsenic 4$p$ electrons. 
In fact, DFT calculations locate the arsenic 4$p$-bands 
not too far from the Fermi energy.
Therefore, a full 3$d$-4$p$ model may provide a satisfactory description 
of the paramagnetic and antiferromagnetic phases of the iron pnictides.

\begin{acknowledgments}
\section{Acknowledgments}
We thank A.~Liebsch for providing us with the DMFT data in Fig.~\ref{fig1}.
\end{acknowledgments}

\bibliographystyle{unsrt}
\bibliography{pnictide}

\begin{thebibliography}{10}

\bibitem{magmom1}
C.~de~la~Cruz~et al.
\newblock {\em Nature~(London)}, 453:899, 2008.

\bibitem{magmom2}
N.~Qureshi et~al.
\newblock {\em Phys.~Rev.~B}, 82:184521, 2010.

\bibitem{magmom3}
H.~F.~Li et~al.
\newblock {\em Phys.~Rev.~B}, 82:064409, 2010.

\bibitem{magmomdft1}
I.~I.~Mazin et~al.
\newblock {\em Phys.~Rev.~B}, 78:085104, 2008.

\bibitem{magmomdft2}
Y.-Z. Zhang, I.~Opahle, H.~O. Jeschke, and R.~Valent{\'i}.
\newblock {\em Phys.~Rev.~B}, 81:094505, 2010.

\bibitem{ishida2010}
H.~Ishida and A.~Liebsch.
\newblock {\em Phys.~Rev.~B}, 81:054513, 2010.

\bibitem{buenemann2003}
J.~B{\"u}nemann et~al.
\newblock {\em Europhys. Lett.}, 61:667, 2003.

\bibitem{hofmann2009}
A.~Hofmann et~al.
\newblock {\em Phys.~Rev.~Lett.}, 102:187204, 2009.

\bibitem{chinesen}
G.~T.~Wang et~al.
\newblock {\em Phys.~Rev.~Lett.}, 104:047002, 2010.

\bibitem{graser2009}
S.~Graser, T.~Maier, P.~Hirschfeld, and D.~Scalapino.
\newblock {\em New Journal of Physics}, 11:025016, 2009.

\bibitem{sugano1970}
S.~Sugano, Y.~Tanabe, and H.~Kamimura.
\newblock {\em {Multiplets of Transition-Metal Ions in Crystals}}.
\newblock Pure and Applied Physics 33, Academic Press, {New York}, 1970.

\bibitem{buenemann1998}
J.~B{\"u}nemann, W.~Weber, and F.~Gebhard.
\newblock {\em Phys.~Rev.~B}, 57:6896, 1998.

\bibitem{buenemann2005}
J.~B{\"u}nemann, F.~Gebhard, and W.~Weber.
\newblock In A.~Narli-kar, editor, {\em Frontiers in Magnetic Materials}.
  Springer, Berlin, 2005.

\bibitem{buenemann2003b}
J.~B{\"u}nemann, F.~Gebhard, and R.~Thul.
\newblock {\em Phys.~Rev.~B}, 67:75103, 2003.

\bibitem{note2}
A.\ Liebsch, private communication, 2010.

\bibitem{lda+u}
F.~Cricchio, O.~Gr{\r{a}}n{\"a}s, and L.~Nordstr{\"o}m.
\newblock {\em Phys.~Rev.~B}, 81:140403, 2010.

\bibitem{buenemann2007b}
J.~B{\"u}nemann, K.~J\'avorne-Radn\'oczi, P.~Fazekas, and F.~Gebhard.
\newblock {\em J. Phys. Cond. Matt}, 19:326217, 2007.

\bibitem{zhou2010}
S.~Zhou and Z.~Wang.
\newblock {\em Phys. Rev. Lett.}, 105:096401, 2010.

\end{thebibliography}

 \end{document}